\documentstyle[11pt]{article}

 \title{ Subtle interface magnetism of Fe/Au multilayers}

\author{ M. B\"ohm{\thanks{Based on the diploma thesis of M. B\"ohm, Regensburg
1997, $^2$ corresponding author, fax
(0049) 941 943 4544, e-mail krey@rphs1.physik.uni-regensburg.de }} \,\, and U.~Krey$^2$
 \\
  Institut f\"ur Physik II, Universit\"at Regensburg,
  Universit\"atsstr.~31,\\ 93040 Regensburg, Germany}

  \date{  September 14, 1998, revised version, accepted by JMMM }

\input epsf
\begin{document}

\large
\maketitle
{\centerline{(Received ................. 1998)}}

\begin{abstract}

By {\it ab initio} LMTO calculations  in atomic-sphere approximation we have
studied the interlayer exchange coupling between Fe films separated by Au
spacers in infinite Fe/Au multilayers with (001) interface orientation. We
also performed detailed calculations of the magnetization and charge
profiles across the system. We find an enhancement of the Fe moments at the
interface, which amount to $\approx 2.8$ $\mu_B$ instead of the bulk value
of $2.2$
$\mu_B$, and we also find a slight magnetic polarization of the order of 0.01
$\mu_B$ at the Au interface layers. These results do not depend
sensitively on the lattice constants
assumed in the calculation.
 When we try to optimize our results with respect to the ratio
$R_{Fe}/R_{Au}$ of the Wigner-Seitz spheres for the mutual components, we
often find the optimum near  {charge neutrality of the interface
monolayers}\,: However, this is not always the case, and usually the sign and
the magnitude of the exchange coupling depend sensitively on the optimal
choice of the above-mentioned ratio. In particular, for
Fe$_5$/Au$_3$ we find an anomaly with a large dipole moment at the interface
and a related anomaly of the exchange coupling.
\end{abstract}
\vspace*{0.5cm}
\section{Introduction}
It is well known meanwhile that in metallic magnetic  multilayers with
 ferromagnetic  films
(e.g. Fe, Co,...) separated by non-magnetic or antiferromagnetic 
spacers ( e.g.~Cu, Cr,...) there exists a pronounced indirect exchange coupling
between the magnetic layers, mediated by the coherent tunneling of the
electrons through the spacer (\cite{Gruenberg,Fert,Parkin}). This coupling is
explained by RKKY-like or electronic Fabry-Perot-like interference theories (e.g.
\cite{Bruno,Bruno2}) or  by theories
stressing the (partial) confinement of certain kinds of the  electrons
(e.g. \cite{Mathon}) in specific layers, or by {\it ab initio} calculations (e.g.
\cite{Sticht,Dederichs,Krompiewski}). Here we have applied an {\it ab initio}
calculation to (001)-Fe$_n$/Au$_m$ multilayers, which are all periodically
continued in the direction perpendicular to the layers, with the purpose to
study in detail not only the above-mentioned exchange coupling, but also to
monitor the changes of the magnetic moment and charge density profiles near
the interface. It was also our intention to see, how the results depend on
details of the computation
 concerning the atomic radii in an atomic-sphere
approximation (ASA) for a binary system.
In this respect, our results turn out to be interesting in itself, and
often they do not depend sensitively on the above-mentioned details.
However, there are exceptions, which we point out, and which show that
sometimes the standard ab-initio calculations, where the atomic positions at
the interface are fixed, are questionable in the present context,
 i.e. one should expect
considerable interface relaxations or even  reconstructions, which
are accompanied by changes in the magnetic coupling {\it et vice versa.} 

\section{The formalism}
We use the LMTO (Linearized Muffin-Tin Orbital)
 program of O.K. Anderson and coworkers in its non-relativistic version and in
the atomic sphere approximation (ASA), however with the so-called 'combined
corrections' and the accurate k-space summation, \cite {Anderson,Bloechl}.
(In all our calculations, the number of k-points was 8000 = 20$\times$
20$\times$ 20).

  The 'combined corrections' take into account (i) the states of higher
angular momentum, e.g.~the f-states, which are otherwise neglected, as
usual, in our LMTO calculations for Fe and Au, and (ii) at the same time
they take into account the fact that the ASA spheres are, on one hand, {\it
overlapping} in parts of the space, whereas on the other hand there remain
'interstitial' regions, which are completely outside the spheres: For
elemental metals the 'combined corrections' reduce the corresponding
mistakes efficiently, \cite{Anderson,Bloechl};
however for the present binary multilayer systems these mistakes may still
belong to the main weaknesses of the ASA
(see below, and \cite{REM1}).

Concerning the structure, we use a hard-core model for Au grown on bcc-Fe
(a=2.78 $\AA$), or sometimes, if explicitly stated, also Fe grown on fcc Au,
with (001) interfaces (fcc nearest-neighbour Au-Au distance: 2.88 $\AA$). As
usual, we assume that the hard-core diameters of the atoms are identical
with the nearest-neigbour distances in both structures). These simplified
structural models for Fe$_n$/Au$_m$ multilayers seem justified
by the thorough studies of Fullerton {\it et al.}, \cite{Fullerton}.

 Our calculation is fully self-consistent concerning charge
densities, spin densities, and energies (with the usual contributions from
the kinetic energy, attraction of the electrons by the nuclei, Madelung
energies from atoms with different nuclear charge, Coulomb repulsion
 and exchange-correlation energy of the electrons), see \cite{Anderson},
 and as in our preceding calculations,
\cite {Sticht,Dederichs,Krompiewski,Suess},  the  standard local spin density
approximation (LSDA) of  van Barth and Hedin  has been used, \cite{Barth}. 

In the atomic sphere approximation for a two-component multilayer
 system, there is however the
following freedom: The 'average sphere radius' $W = [3V/(4\pi N)]^{1/3}$ is of
course fixed for a given multilayer, where
$N$ is the total number of atoms in an elementary cell, and $V$
($=n_1V_1+n_2V_2$) its volume, however the ratio
\begin{equation}
R_1/W = \{N
/[n_1+n_2 (V_2/V_1)]\}^{1/3}\end{equation} is not.
 Here $n_1$ resp.~$n_2$ are the numbers of Fe resp.~Au atoms in the
elementary cell, and $V_1$ resp.~$V_2$ are the
corresponding  atomic volumes, i.e.~$V_i=4\pi R_i^3/3$.

 In our former calculations, we have always taken for $V_1$ and $V_2$ those
values, which these quantities have in the  elemental metals, namely
$V_{Fe}=11.82$ $\AA^3$ and $V_{Au}=16.98$ $\AA^3$, which
corresponds to $R_1/W\approx 0.94$ for $n_{Fe}=n_{Au}=1$. Another
choice would be $R_i/W=1$. However here we consider
$R_1/W$ as a {\it variational parameter} for our calculation, and thus we
present results below, where $R/W$ (=$R_1/W = R_{Fe}/W$) is varied between
$\approx 0.94$ and $1$.  We stress at this place that every variation of
$R_1$ must of course be accompanied by a well-defined change of $R_2$, such
that $n_1R_1^3+n_2R_2^3$ remains fixed. Also all atomic positions remain
fixed in our approach. 
  
\section{Results} In Fig.~1, we present results for
Fe$_2$/Au$_2$-multilayers, always with the above-mentioned
(001)-orientation. The figure contains four parts, namely

(i)  in the upper-left quadrant  the total charge $Q$ ($=Q_{tot}$ in the figures),
which is contained, according to our calculation, in a Fe
ASA-sphere at the interface layer, is presented in units of the electronic
charge $q_e$. The quantity $Q$ thus measures the local deviation from charge
neutrality at the atom considered. (Positive $Q[q_e]$  means an excess
of electronic charge, compared with the neutral atom.)\,\, One can see from
Fig.~1 that the interface is electrically neutral at $(R/W)\approx 0.959\pm
0.001$, whereas for larger (smaller) ratio $R/W$ the Fe interface layer
contains a higher (smaller) amount of electronic charge, as expected. The
dependence of $Q$ on $R/W$ is linear.
 
 (ii) In the lower-left quadrant of Fig.~1, the minimum of the
total energy per antiferromagnetic elementary cell of 8 atoms
appears also at the same value $R/W\approx 0.959\pm 0.001$, and
  to the
accuracy of the drawing the results cannot be distinguished for
 mutually parallel resp.~antiparallel alignment of the 
ferromagnetic layers.

(iii)  But on the upper-right quadrant of Fig.~1 the energy difference
$\Delta E :=E^{\uparrow \uparrow}-E^{\uparrow \downarrow}$ per
antiferromagnetic unit cell (8 atoms) is presented, and one sees that the
{\it ferromagnetic state} is energetically slightly favoured at the
above-mentioned value of $R/W\approx 0.959\pm 0.001$, whereas for $R/W
\,\widetilde{<}\,\,
0.95$ and $R/W\, \widetilde{>}\,\, 0.98$ one would predict 
a different mutual orientation, namely the {antiferromagnetic} one.

(iv)
Finally the lower-right quadrant of Fig.~1 shows the Fe moments
 at the interface, which vary
only slightly between $2.7$ and $2.9$ $\mu_B$, when $R/W$ increases from
0.94 to 1, and at the 'optimal value' of $R/W\approx 0.959$ the moment is
$\approx 2.775\pm 0.005$ $\mu_B$, both for mutual parallel  resp.~antiparallel
alignment. 

These results were calculated for Au grown on Fe, however similar results
 are also obtained for the slightly different structure corresponding to Fe
grown on Au. Since these results do hardly differ, they are not plotted
here.

As a consequence, to get relevant results for the composition n=m=2 of our
multilayer it {\it appears} that, to a first approximation, one should
simply take that value of
$R/W$, where one has charge neutrality at every atom.
However, this simple recipe does apparently not work as a general rule:
 E.g.~for $n=m=1$ we find that the minimum of the total energy is taken for
a higher value of $R/W$ slightly below $0.98$ with $Q[q_e]$ as large as
$\approx 0.125$, whereas charge neutrality
would again happen near $R/W\approx 0.96$. Also for
$n=5,\, m=3$  we obtain pronounced deviations from the
above-mentioned 'postulate' (see below); so, to our experience, it would be
unreasonable to take local charge neutrality as an unchecked 'natural
apriori-approximation'.

In fact, in Fig.~2a we present our results for the Fe$_5$/Au$_3$ multilayer,
again for Au grown on Fe. The minimum of the total energies per
antiferromagnetic elementary cell of 16 atoms, both for mutually parallel
and for mutually antiparallel orientation of the Fe magnetizations of
subsequent Fe sandwiches, happens again for $R/W\approx 0.96$ in this case,
but now for this value there is a large charge transfer of $Q\approx-0.05$
$q_e$ at the Fe interface layers, i.e.~the Fe rsp.~Au layers at the
interface carry a positive (resp.~negative) charge of $\mp 0.05$ $q_e$ per
atom.
 In contrast, charge neutrality at the interface would now happen at a
significantly higher value of $R/W\approx 0.973$. At this higher value, the
exchange interaction is {\it antiferromagnetic} and relatively small (i.e.~the energy
 difference $\Delta E$, upper-right
quadrant of Fig.~2a, is positive, of the order of 0.0001 Ry),
 whereas at the sharp minimum of the total energy in Fig.~2, i.e.~at
$R/W\approx 0.96$, the exchange energy is definitely ferromagnetic and one
order of magnitude larger, namely $\Delta E
:=E^{\uparrow\uparrow}-E^{\uparrow\downarrow}\approx - 0.003$ Ry for our
antiferromagnetic unit cell of 16 atoms. 

We have repeated these subtle calculations for a slightly modified
structural model, corresponding now to Fe grown on Au, see Fig.~2b, and on
this occasion we have produced data for an additional point of $R/W$ just
above the minimum at $R/W\approx 0.96$.  From this additional calculation it
seems that the pronounced minimum at
$R/W\approx 0.96$, with $\Delta E\approx -0.003$ Ry, is even much steeper
than expected from Fig.2a, and it seems that here some kind of resonance
phenomenon happens which is beyond the simplifications made in the LMTO-ASA
method, see below.

In fact, if one plots the charge $Q$ per atom for the eight different layers
of our periodically continued and ferromagnetically polarized Fe$_5$/Au$_3$
system against the layer index, one gets the results presented in Fig.~3.
 Here the solid circles are for the
 energetically stable configuration with $R/W\approx 0.96$, where a
strong negative dipole moment (the Fe sphere has a positive charge,
since $q_e$ is negative) at the interface from layer five (Fe) to layer 6
(Au) is observed. In contrast, for the  above-mentioned configuration
  with $R/W\approx 0.973$, (the open circles), the interface dipole moment
  is reduced by two-thirds in magnitude, and is, moreover, {\it inverted in
  sign}. This kind of behaviour gives rise to speculations that this
approximate multivaluedness may be resolved by some kind of {interface
  reconstruction} where the large dipole moments are reduced to quadrupole
  moments, or by some kind of interdiffusion, or by the formation of an
  interface alloy {\cite{Bluegel}: These possibilities, to be studied for
  systems as large as the present one, are beyond our present computational
  abilities.  To our opinion, they demand an extremely accurate treatment by
  a full-potential method, i.e.~beyond the ASA, beyond LDA, for more general
  structures, and perhaps also with non-collinear spin configurations. 

Here it should of course be stressed that in Fe$_5$/Au$_3$ one has actually
three non-equivalent Fe layers and two non-equivalent Au layers, so that our
approach with only one variational parameters should at least in principle
be replaced by an approximation with four variational parameters,
($R_{Fe}$)$_i$, with i=1,2,3, and $\alpha :=(R_{Au}$)$_1$/($R_{Au}$)$_2$
(see the footnote \cite{REM2}), which is however again beyond our
computational capabilities. Instead, we restricted ourselves to the case
where the first three variational parameters, ($R_{Fe}$)$_i$, are equal and
the fourth parameter,
$\alpha$, is 1. Our above-mentioned parameter value of $R/W\approx 0.973$,
corresponding to the flat local minimum of the energy difference, might thus
in fact be closer to the (four-dimensional) global energy minimum than the
above-mentioned value of
$R/W\approx 0.96$, where according to our (one-dimensional) variational
approximation the minimum is situated.

Interestingly, the anomalies seen in Fig.~2 apparently do not show up in in
other quantities: In particular,
  when plotting the optimal radii $R_{Fe}$ and $R_{Au}$ determined in
  our calculation for the four systems (1) Fe$_1$/Au$_1$,
  (2) Fe$_2$/Au$_1$, (3) Fe$_2$/Au$_2$, and (4) Fe$_5$/Au$_3$, we find the
  results presented in Fig.~4.
From this figure one concludes that the
  optimal value of $R_{Fe}$ in our calculations practically does not change,
  and it is only $R_{Au}$ that varies. If one extrapolates this result,
  i.e.~the approximate constancy of the 'optimal' $R_{Fe}$, also to other
  compositions, it may be quite
useful, since with $W$ and $R_{Fe}$ also the optimal
  values of $R_{Au}$ would be known.

In Fig.~5 we also plot profile-functions of the magnetic moments. From
the figure one can not only see that the iron moments near the interface
are enhanced, as mentioned above, but one can also see that the Au
atoms, too, become slightly polarized at the interface, to the order of 0.01
$\mu_B$. This Au polarization is parallel to that of Fe, whereas
at the second Au layer, it is antiparallel, but still much smaller.
Experimentally such small moments can be measured by X-ray dichroism,
\cite{Schuetz}, and the enhanced Fe  moments at an (001) interface to Au
 have been found
by experimental work in our department, \cite{Hoepfl}.

 In Fig.~6a,b and Fig.~7 we finally plot results for the interlayer
 exchange coupling $J$ as a function of the Au thickness $x$ (Fig.~6) and
of the Fe thickness, (Fig.~7), for different systems.
 Obviously it is necessary to take the optimized value of $R/W$, and not the
'old' one obtained with the ratio $R_{Fe}/R_{Au}$ taken from bulk
calculations; i.e.~from Fig.~6 and Fig.~7 we find that the computational
results for the exchange interaction are astonishingly sensitive to the
choice of $R/W$. One could be tempted to extrapolate the 'new' results of
Fig.~6a by a decaying spatially-sinusoidal exchange oscillation of the form $\Delta E
\sim 0.0006\,$ Ry $\times \sin [\pi\cdot (x-2)/3]/(x/3.5)^2$,
 i.e.~with a 'period' of roughly 6 Au monolayers. This would look reasonable in
view of the expected asymptotic behaviour, \cite{Bruno,Bruno2}; however
actually, in Fig.~6, one is still very far from the asymptotic regime; so
this extrapolation should not be taken serious, although from $x=2$ to $x=5$
it fits the data quite well. At
the same time, from Fig.~6b it seems that the 'unnatural behaviour',
particularly with the drastic change observed between Fe$_3$Au$_4$ and
Fe$_3$Au$_5$ with the 'old' parameters, looks much smoother now, and more
reasonable, with the new optimized parameters, in agreement with the smooth
behaviour already mentioned in connection with Fig.~4. Finally in Fig.~7 it
is obvious that the dependence on $n_{Fe}$ is more drastic for the
Fe$_n$/Au$_1$ system than for Fe$_n$/Au$_2$, which is reasonable, since also
the deviations from local charge neutrality are much larger in the
first-mentioned case:

Actually, at the optimal value of the variational parameter $R/W$,  we have
observed the most pronounced deviations from local charge neutrality for
Fe$_1$/Au$_1$ multilayers, whereas for Fe$_2$/Au$_2$ we had charge
neutrality at the optimum. The difference is plausible on symmetry
reasons\,:

 In the first-mentioned case, an Fe atom has two Au neighbours at the
right-hand rsp.~left-hand side, say, and charge transfer from these
neighbours into the overlap region of the Fe ASA-sphere sums up to a
non-zero value. In contrast, in the second case, an Fe-atom has one Au
neighbour, say, to the left, and a Fe neighbour to the right; if there is
now a charge transfer in the overlap region from Au to Fe, i.e.~from the
left, say, due to a {\it reduction}
$\delta R_{Au}<0$, this corresponds to an {\it enhancement}  $\delta
R_{Fe}>0$ for the right neighbour. I.e.~at the overlapping region to the right
 the charge transfer from the Fe neighbour will probably have opposite sign
to that one observed at the overlapping region to the left. This will lead
to largely compensating transfers, and to a correspondingly small result for
$|Q|$. Therefore, even if for Fe$_2$/Au$_2$ multilayers the single ASA
spheres are essentially charge-neutral, they will probably carry a large
non-trivial charge-density, e.g.~positive rsp.~negative, near the left-hand
rsp.~right-hand overlap regions of the ASA spheres; whereas for
Fe$_1$/Au$_1$ these regions will carry charges of the same sign leading to
large values of $|Q|$. In the overlapping regions and nearby, the magnitudes
of the local charge transfer should be of the same order in both cases.
\section{Conclusions}
We have calculated   the spatial variation of the
charge-density, the profile-functions of the local moments, and the exchange
coupling energy between successive Fe films, for
 (001)-Fe$_n$/Au$_m$ multilayers, by a
LMTO calculation in ASA and LSDA approximations, and observed subtle
behaviour: Varying the ratio of the
Wigner-Seitz radii $R_{Fe}/R_{Au}$ for given value of the 'average
Wigner-Seitz radius $W$', see eq.~(1), we found that often -- but not always
-- the best values for the total energy (and as a consequence also for the
interlayer exchange couplings of the Fe layers across the Au spacer) is
obtained near 'charge neutrality' of the interfaces. In particular for
Fe$_1$/Au$_1$ multilayers and Fe$_5$/Au$_3$ multilayers this was {\it not} the
case. For the last-mentioned system a subtle kind of 'resonant behaviour' as
a function of $R/W$ appeared near the value
$R/W=0.96$, and as a consequence there may be in this case a strong interface
reconstruction, or other possibilities like an interface alloy, or
noncollinear states, which is beyond the approximations and limitations of
the present approach.
In fact, it is our belief that
with the present paper, which first originated as a case study, we have also
made obvious that such subtle problems as the present one, or even more
subtle problems as the just mentioned 'other possibilities', should better
be treated -- if possible -- by a full-potential formalism, and possibly
with non-collinear spin states. At the same time there is  a demand
for conclusive experiments.

  \subsection *{Acknowledgements} Discussions with A.~Moser, F.~S\"uss,
   S.~Bl\"ugel,
   R.~Wiesendanger, C.~Demangeat and G.~Bayreuther are gratefully
  acknowledged.
   We
  also thank the Computing Center of the University of
  Regensburg and the LRZ in Munich 
   for  computing time.

\newpage
\centerline{\bf{Figure captions}}
{\noindent{ {\underline{\bf{Fig.1:}} (i) The charge $Q_{tot}$ in an
interface Fe ASA sphere (always $=Q$
in the text) in units of the (negative) electronic charge $q_e$, (ii)
 the total
energy $E_{tot}$ ($=E$ in the text) of the antiferromagnetic unit
cell containing eight atoms, (iii) the energy difference $\Delta E
=E^{\uparrow\uparrow}-E^{\uparrow\downarrow}$, and (iv) the magnetic moment
of the interface Fe ASA sphere are plotted against the ratio $R/W$, where $R$
is the radius of the Fe sphere used in our LMTO-ASA calculation for
$Fe_2/Au_2$ multilayers, while $W$ is fixed by the equation $4\pi
W^3/3=N/V$, where $V$ is the volume and $N$ the number of atoms of our
multilayer. The 'hard-core' structural model has been produced by growing Au
on bcc-Fe, with (001)-interfaces. Our multilayers are always periodically
continued in the direction perpendicular to the interfaces.}

{\noindent{\underline{\bf{Fig.2a:}} The same as in Fig.1, but for
$Fe_5/Au_3$ multilayers.}

{\noindent{\underline{\bf{Fig.2b:}} The same as in Fig.2a, but for Fe grown
on Au. In the lower right figure, the solid and dotted lines, respectively,
remind to the slightly different results in Fig.2a. The lines are a guide to
the eye only, and Fig.2b, part (iii) for $\Delta E$, shows that actually
near the resonance more points are needed.}

{\noindent{\underline{\bf{Fig.3:}}
 For the stable state at $R/W\approx 0.96$ in Fig.2a (full circles),
and for the different state at $R/W\approx 0.975$ (open  circles), the
charge $Q_{tot}$ contained in the respective  atomic  spheres is
plotted against the layer index. Layers 1, ..., 5 correspond to Fe, the rest
to Au. Note the drastic change of sign and magnitude at the layers 5 and 6.}

{\noindent{\underline{\bf{Fig.4:}} For the cases 1$\hat = Fe_1/Au_1$, 2$\hat =
Fe_2/Au_1$, 3$\hat
=Fe_2/Au_2$, and 4$\hat = Fe_5/Au_3$, the radius $R_{min}$ corresponding
 to the absolute minimum of $E_{tot}$ (see e.g.~Figs.1--2 for cases 3 and 4)
is plotted for Fe
(filled circles) and Au (open circles).}

{\noindent{\underline{\bf{Fig.5:}} The profiles of the magnetic moments per
ASA sphere across the multilayer are plotted against the layer index for
$Fe_2/Au_3$, $Fe_3/Au_5$,
$Fe_5/Au_1$ and $Fe_5/Au_2$, both for parallel and antiparallel mutual spin
orientation of the Fe sandwiches. The induced Au moments have been enlarged
by a factor of 10. }

{\noindent{\underline{\bf {Fig.6a:}} The energy difference $\Delta E = E^{\uparrow\uparrow}-
E^{\uparrow\downarrow}$ per antiferromagnetic elementary cell is plotted 
against the number $x$ of Au layers for $Fe_2Au_x$ multilayers.
The filled circles are the new results with the optimized ratio of $R/W$,
whereas  the open circles correspond to the ratio $R/W$ obtained from the
bulk values of $R_{Fe}$ and $R_{Au}$.}

{\noindent{\underline{\bf{Fig.6b:}} The same as Fig.6a, but for $Fe_3/Au_x$ multilayers.}

{\noindent{\underline{\bf{Fig.7:}} The same as in Fig.6, but now in both cases the
'new method' is used and the Fe thickness is varied. The 'filled circles' are
for $Fe_x/Au_1$, the 'open circles' for $Fe_x/Au_2$.}

}}
\newpage
  \epsfxsize=12.5cm
  \epsfbox{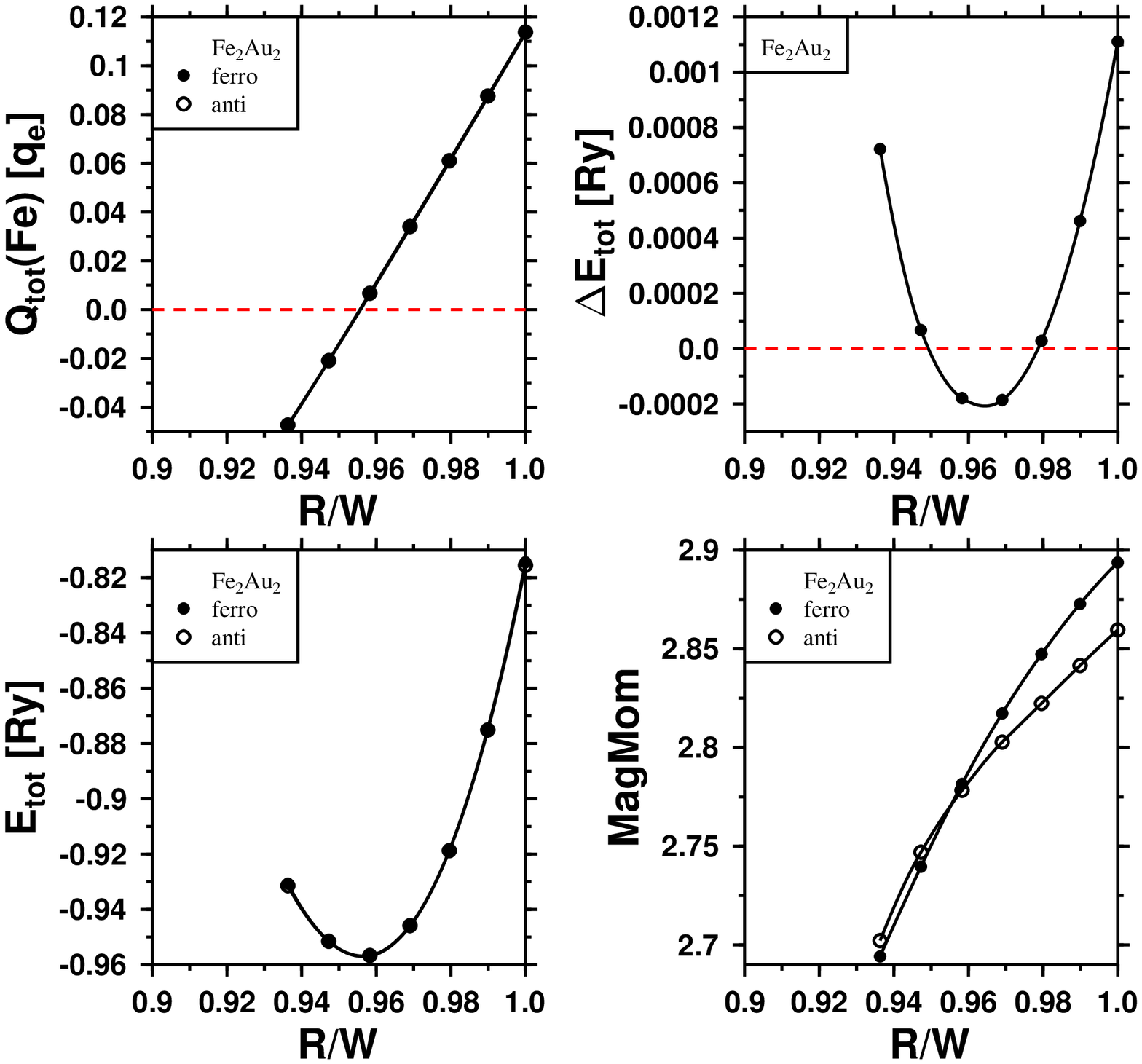}
{Fig.1: (i) The charge $Q_{tot}$  in an interface Fe ASA sphere
 (always $=Q$ in the text)
 in units of the (negative) electronic
charge $q_e$, (ii)
 the total
energy $E_{tot}$ ( $=E$ in the text) 
of the antiferromagnetic unit
cell containing eight atoms, (iii) the energy difference $\Delta E
=E^{\uparrow\uparrow}-E^{\uparrow\downarrow}$, and (iv) the magnetic moment
of the interface Fe ASA sphere are plotted against the ratio $R/W$, where $R$
is the radius of the Fe sphere used in our LMTO-ASA calculation for
$Fe_2/Au_2$ multilayers, while $W$ is fixed by the equation $4\pi
W^3/3=N/V$, where $V$ is the volume and $N$ the number of atoms of our
multilayer. The 'hard core' structural model has been produced by growing Au
on bcc-Fe, with (001)-interfaces.  Our multilayers are always periodically
continued in the direction perpendicular to the interfaces.}
\label{fig1}
\newpage
  \epsfxsize=7.5cm
  \epsfbox{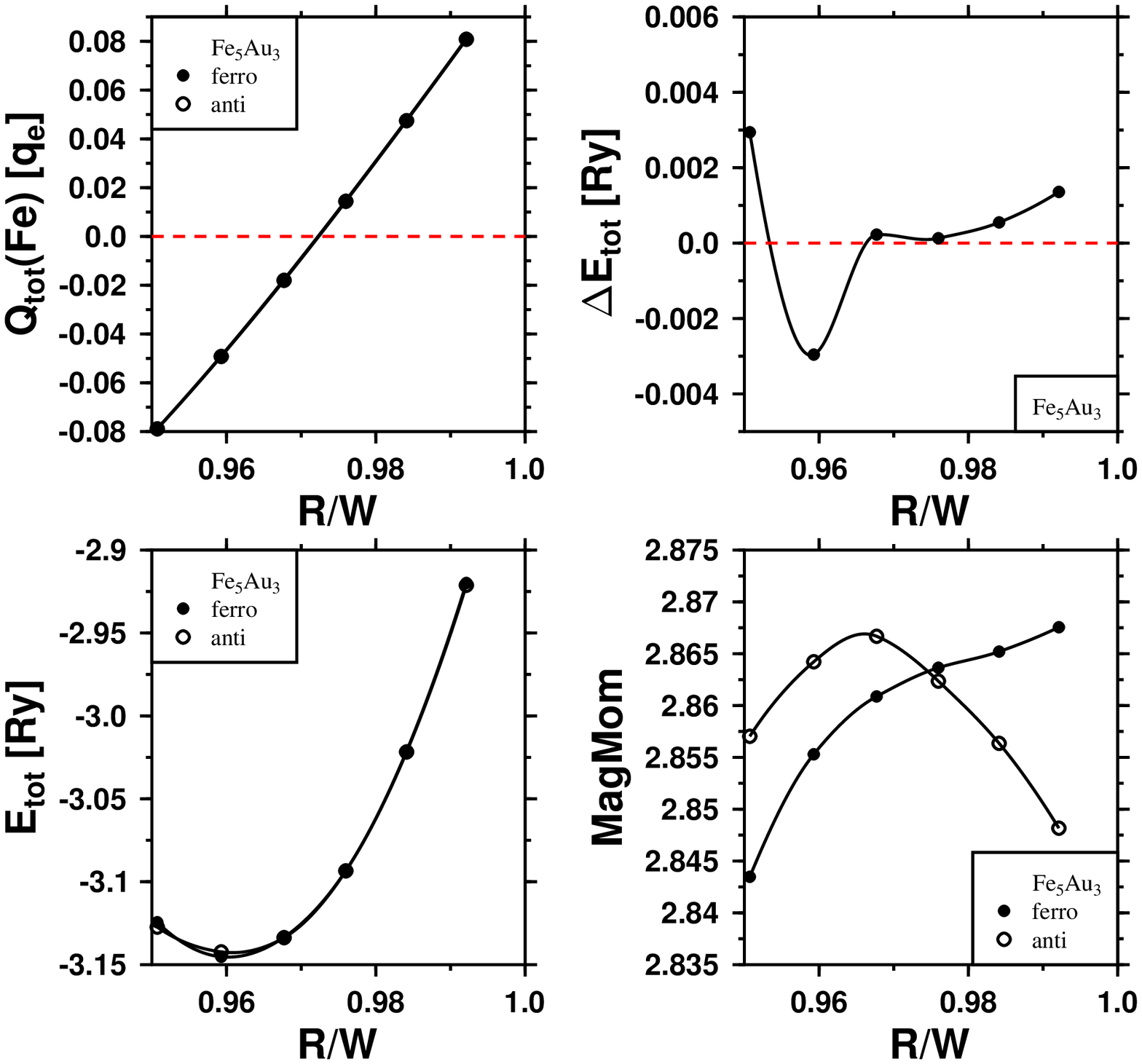}

\noindent{Fig.2a: The same as in Fig.1, but for $Fe_5/Au_3$ multilayers.}

\label{fig2}
  \epsfxsize=7.5cm
\epsfbox{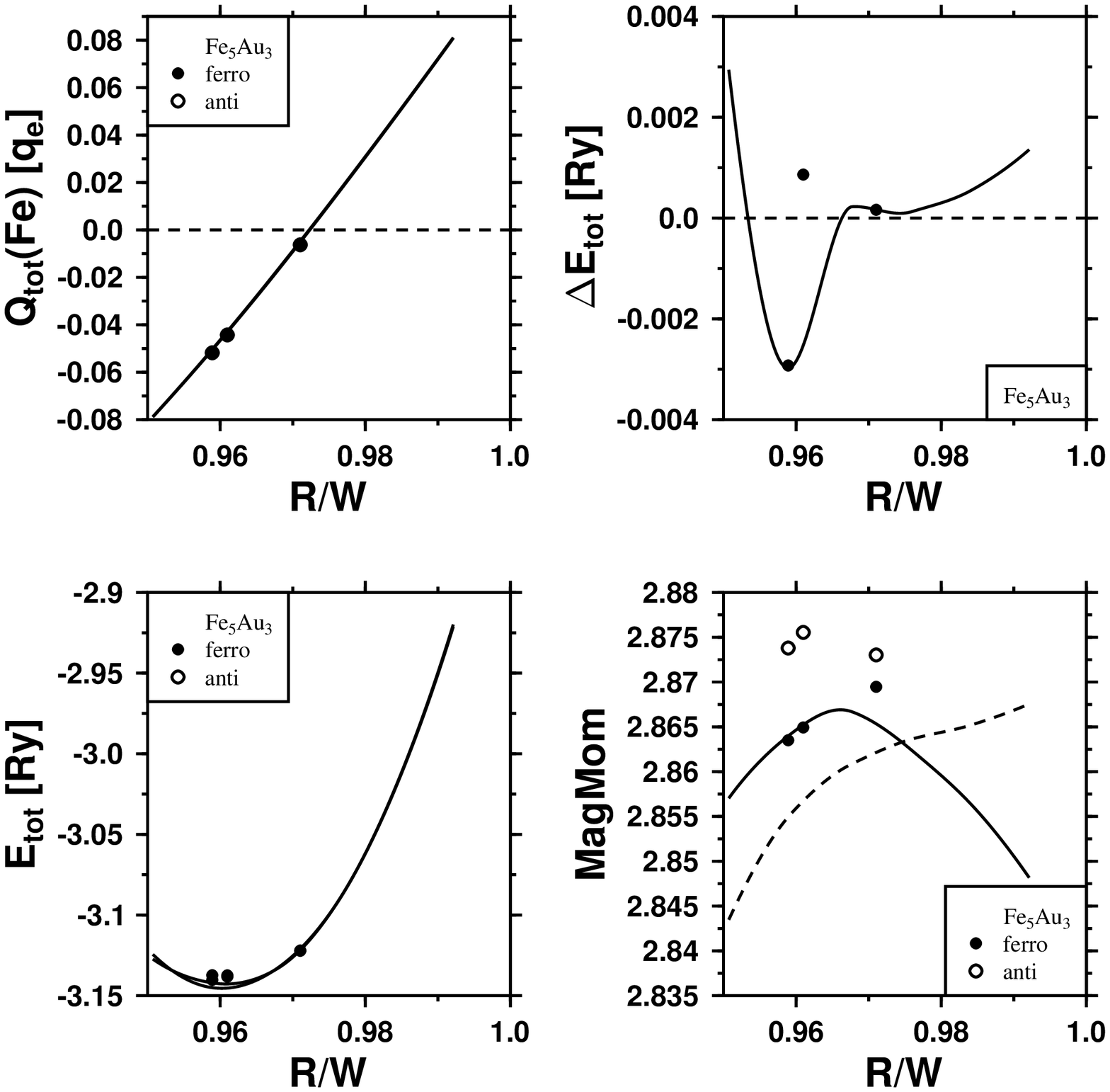}

\noindent{Fig.2b: The same as in Fig.2a, but for Fe grown on Au. In the lower right
figure, the solid and dotted lines, respectively, remind to the slightly
different results in Fig.2a. The lines are a guide to the eye only, and
Fig.2b, part (iii) for $\Delta E$,
shows that actually near the resonance more points are needed.}
\label{fig3}
\newpage
  \epsfxsize=7.0cm
  \epsfbox{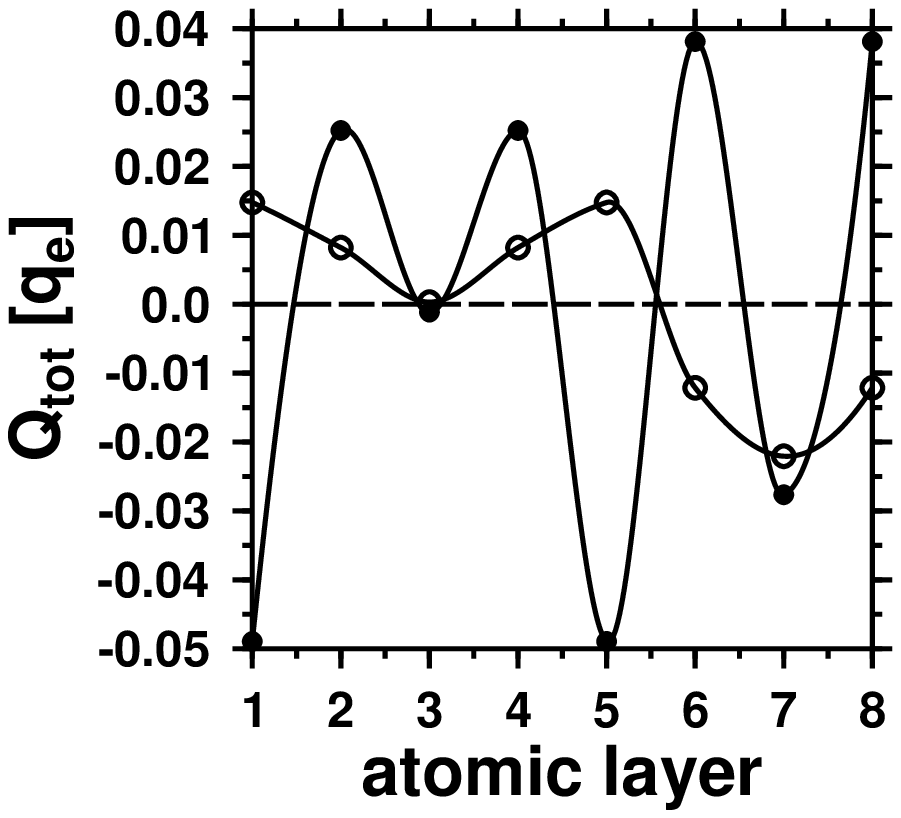}

\noindent{Fig.3: For the stable state at $R/W\approx 0.96$ in Fig.2a (full circles),
and for the different state at $R/W\approx 0.975$ (open  circles), the
charge $Q_{tot}$ contained in the respective  atomic  spheres is
plotted against the layer index. Layers 1, ..., 5 correspond to Fe, the rest
to Au. Note the drastic change of sign and magnitude at the layers 5 and 6.}
\label{fig4}

  \epsfxsize=7.0cm
  \epsfbox{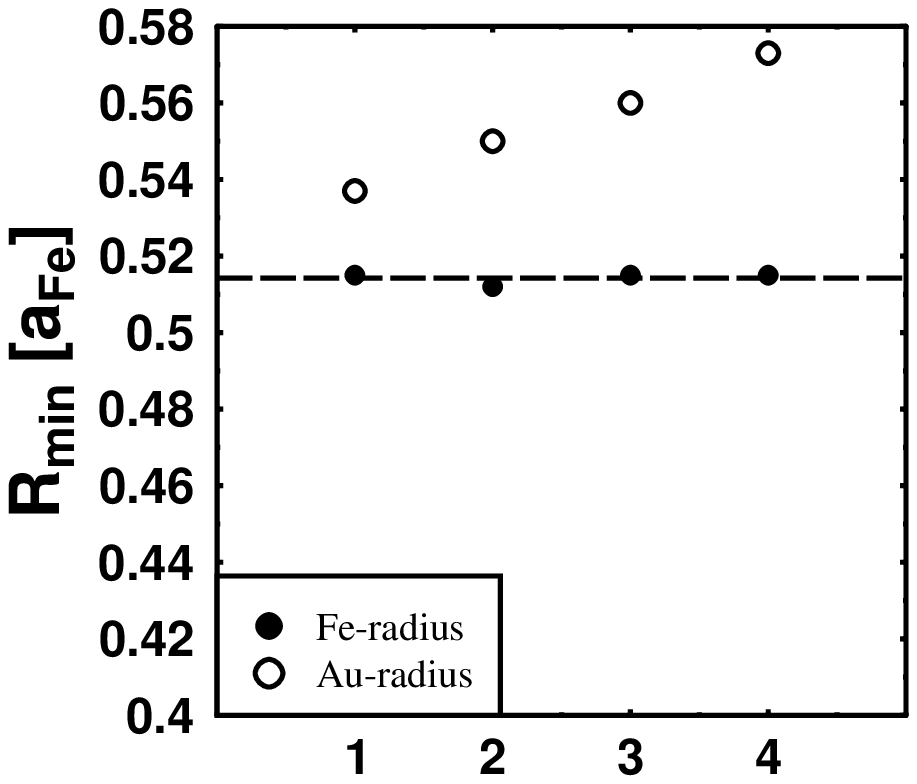}
{Fig.4: For the cases 1$\hat = Fe_1/Au_1$, 2$\hat = Fe_2/Au_1$, 3$\hat
=Fe_2/Au_2$, and 4$\hat = Fe_5/Au_3$, the radius $R_{min}$ corresponding
 to the absolute minimum of $E_{tot}$ (see e.g.~Figs.1--2 for cases 3 and 4)
is  plotted for Fe
(filled circles) and Au (open circles).}
\label{fig5}

\newpage
  \epsfxsize=11.5cm
  \epsfbox{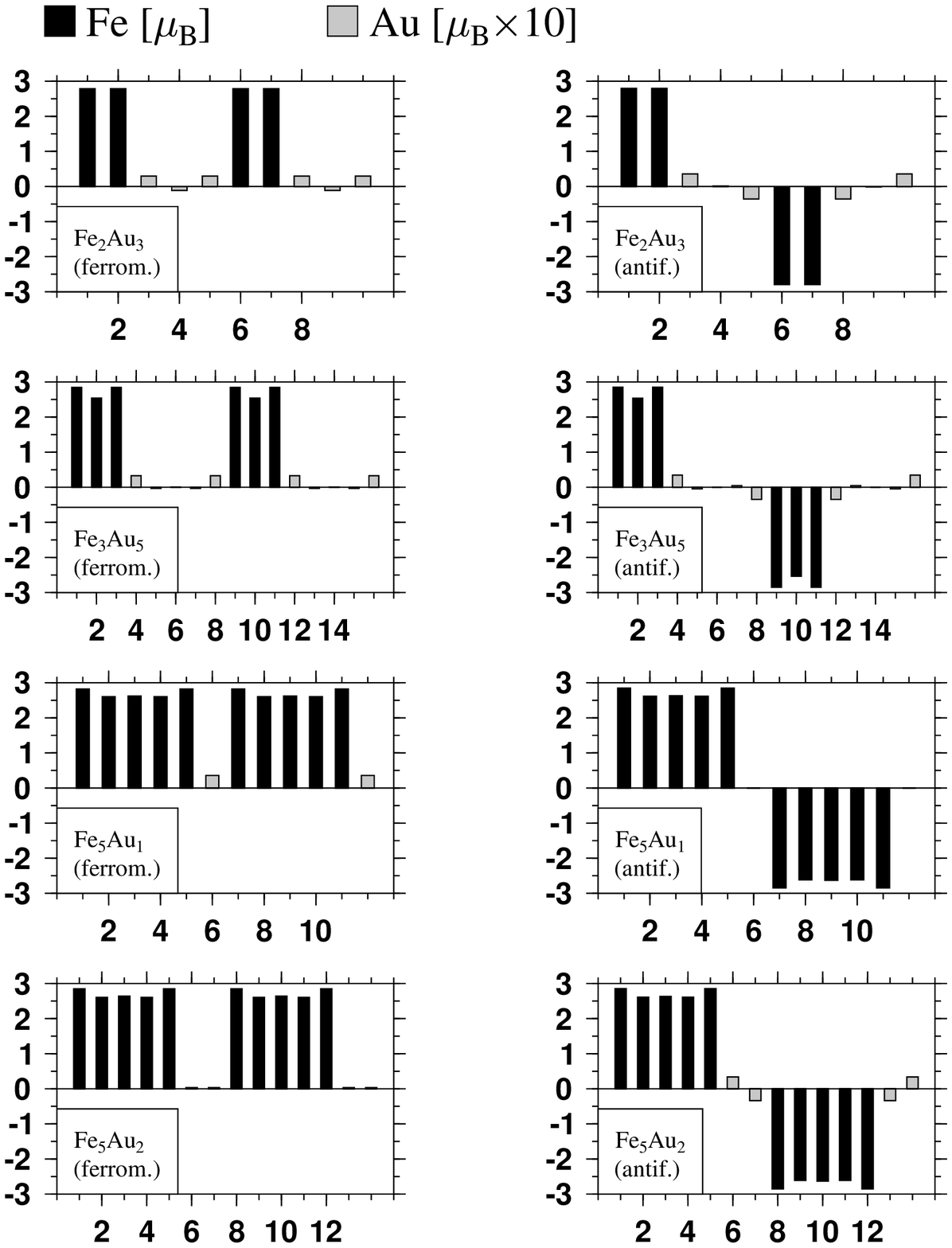}
{Fig.5: The profiles of the magnetic moments per ASA sphere across the 
multilayer are plotted against the layer index for $Fe_2/Au_3$, $Fe_3/Au_5$,
$Fe_5/Au_1$ and $Fe_5/Au_2$, both for parallel and antiparallel mutual spin
orientation of the Fe sandwiches. The induced Au moments have been enlarged
by a factor of 10. }
\label{fig6}

  \epsfxsize=9.5cm
  \epsfbox{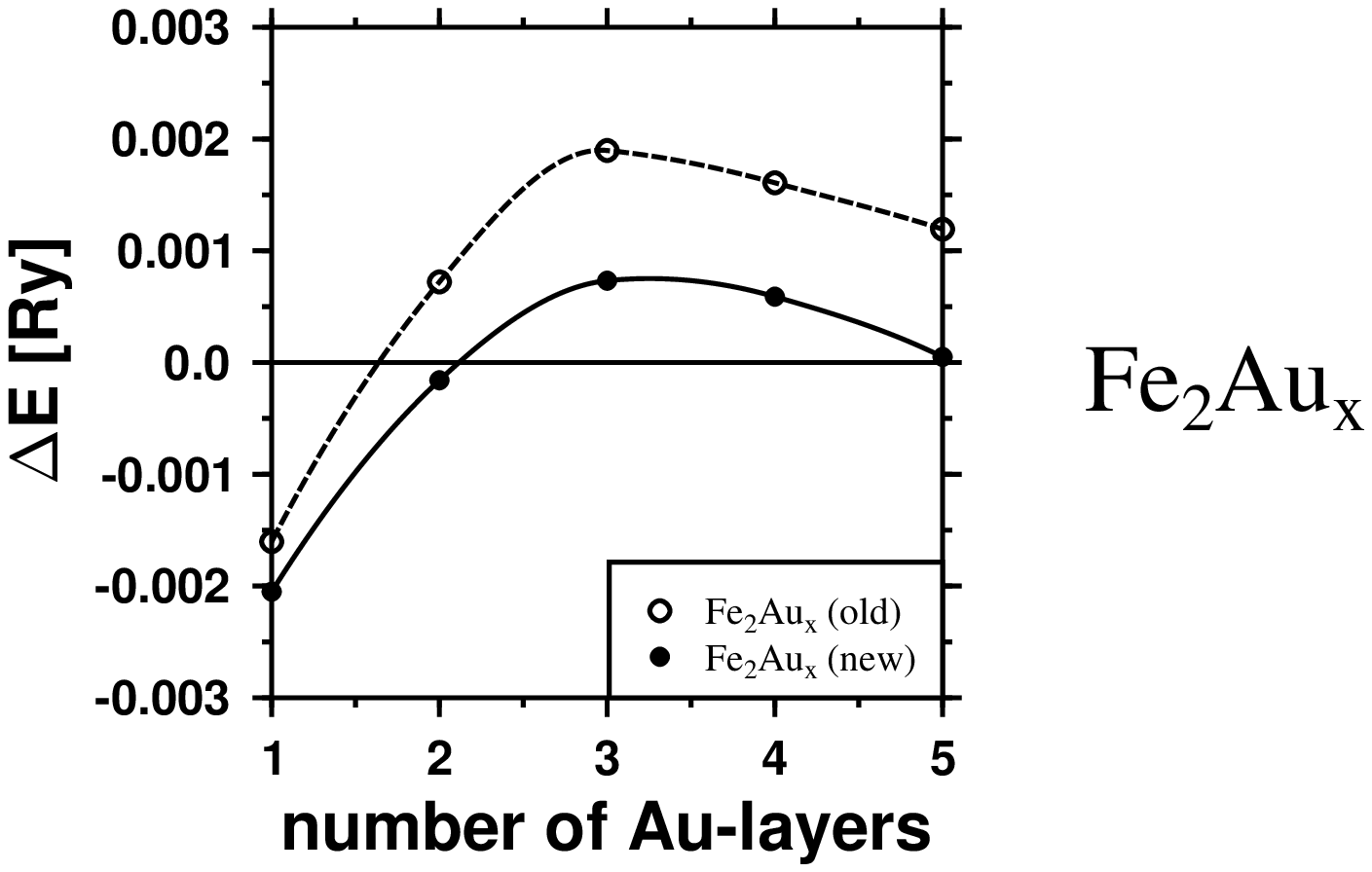}

\noindent{Fig.6a: The energy difference $\Delta E = E^{\uparrow\uparrow}-
E^{\uparrow\downarrow}$ per antiferromagnetic elementary cell is plotted 
against the number $x$ of Au layers for $Fe_2/Au_x$ multilayers.
The filled circles are the new results with the optimized ratio of $R/W$,
whereas  the open circles correspond to the ratio $R/W$ obtained from the
bulk values of $R_{Fe}$ and $R_{Au}$.}

\label{fig7}
  
\epsfxsize=9.5cm
 \epsfbox{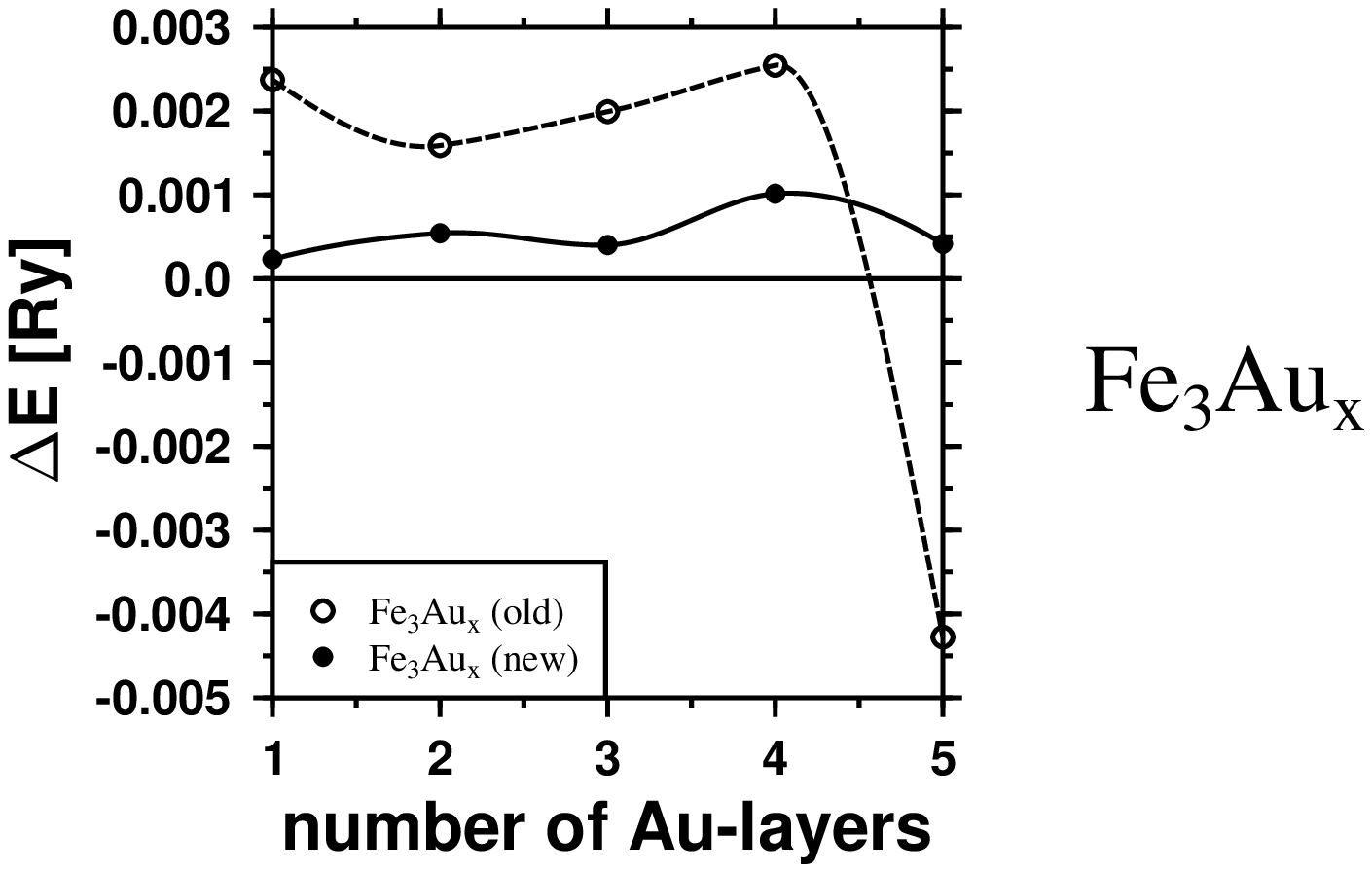}

\noindent{Fig.6b: The same as Fig.6a, but for $Fe_3/Au_x$ multilayers.}
\label{fig8}
\newpage
  \epsfxsize=11.0cm
  \epsfbox{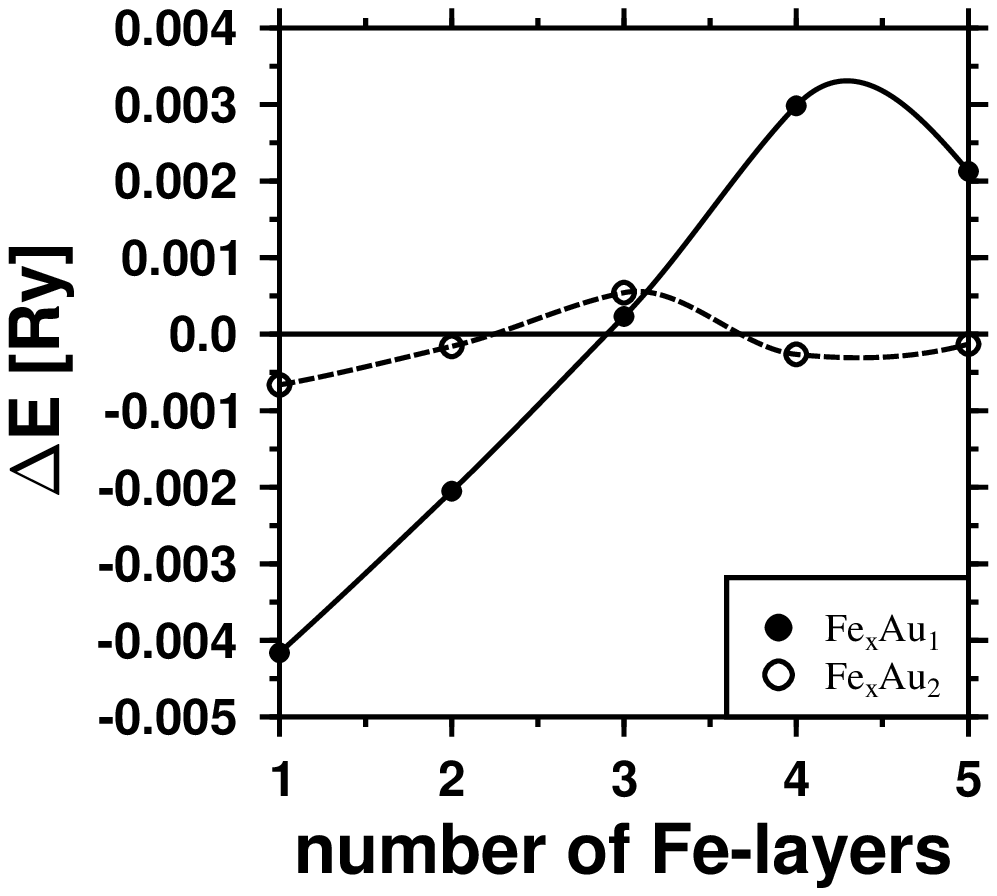}
{Fig.7: The same as in Fig.6, but now in both cases the
'new method' is used and the Fe thickness is varied. The 'filled circles' are
for $Fe_x/Au_1$, the 'open circles' for $Fe_x/Au_2$.}
\label{fig9}

\begin{thebibliography}{99}
\bibitem{Gruenberg} P. Gr\"unberg, R. Schreiber, Y. Pang, M.B. Brodsky and
H. Sowers, Phys. Rev. Lett. 57 (1986) 2442
\bibitem{Fert} M.N. Baibich, J.M. Broto, A. Fert, F. Nguyen van Dau, F.
Petroff, P. Etienne, G. Creuzet, A. Friederich and J. Chazelas,
  Phys. Rev. Lett. 61 (1988) 2472
\bibitem{Parkin} S.S.P. Parkin, N. More and K.P. Roche, Phys. Rev. Lett. 64
(1990) 2304.
\bibitem {Bruno}  P. Bruno and C. Chappert, Phys. Rev. Lett. 67 (1991) 1062
\bibitem{Bruno2} P. Bruno, Phys. Rev. B 52 (1995) 411.
\bibitem{Mathon} D.M.  Edwards , J. Mathon, R. B. Muniz and M.S. Phan, Phys.
Rev. Lett. 67 (1991) 493.
\bibitem{Sticht} F. Herman, J. Sticht and  N. van Schilfgaarde, in: Magnetic
Thin Films, Multilayers and Surfaces, Eds. S.S.P. Parkin et al., Vol 231 of
MRS Symp. Ser., Anaheim, Spring 1991 (Materials Research Society,
Pittsburgh, 1992) p.195.
\bibitem{Dederichs} L. Nordstr\"om, P. Lang, R. Zeller and P.H. Dederichs,
Phys. Rev. B50 (1994) 13058.
\bibitem{Krompiewski} S. Krompiewski, F. S\"uss and U. Krey, Europhys. Lett.
26 (1994) 303.
\bibitem{Anderson} O.K. Anderson, O. Jepsen, D. Gl\"otzl, in:
Highlights of Condensed Matter Theory, LXXXIX Corso di Varenna, Bologna
1985, p. 59. We thank O.K. Andersen and O. Jepsen for the
LMTO package
\bibitem{Bloechl} Peter E. Bl\"ochl, O. Jepsen, O.K. Andersen, Phys. Rev. B 49
(1994) 16223
\bibitem{REM1} The main weakness of the LMTO-ASA method in contrast to
so-called 'Full-Potential' methods is probably the use of spherically
averaged potentials in the ASA spheres, which is especially questionable
near the interfaces.
\bibitem{Fullerton} Eric E.~Fullerton. D. Stoeffler and K.~Ounadjela,
B.~Heinrich and Z.~Celinski, J.A.C.~Bland, Phys.~Rev.~B51 (1995) 6364
\bibitem{Suess} S. Krompiewski, F. S\"uss and U. Krey, J. Magn. Magn. Mater.
164 (1996) L263
\bibitem{Barth} U. van Barth and L. Hedin, J. Phys. C5 (1972) 1629
\bibitem{Bluegel} S. Bl\"ugel, Appl. Phys. A 63 (1996) 595
\bibitem{REM2} A calculation with more than one variational parameter $R_i$
would be similar in spirit as that of \cite{Fullerton}. There the authors
performed a variational calculation with two resp.~three variational
parameters for (001)-Fe$_3$/Pd$_1$ and Fe$_3$/Pd$_2$ systems, using an
ASW-ASA code. However, in contrast to the present paper, they only
considered the case of mutually parallel orientation of the magnetic
moments. But in their calculation they changed the interlayer distances in
accordance with the variations $\delta R_i$, whereas in our calculation all
atomic positions are kept fixed. In fact, in \cite{Fullerton} the authors
observed large effects only due to the Pd atoms. In particular, in a
study of Fe/Au bilayers and Fe/Pd/Au trilayers, grown by molecular epitaxy
on Ag(001), it was also found in \cite{Fullerton} that the Fe and Au layers
were well represented by their bulk structure, in agreement with our
structural assumptions.
\bibitem{Schuetz} G. Sch\"utz, M. Kn\"ulle, H. Ebert, Physica Scripta T49
(1993) 302.
\bibitem{Hoepfl} T. H\"opfl, Diploma thesis, University of Regensburg 1996,
unpublished; M. Brockmann, L. Pfau, G. Lugert and G. Bayreuther, Mat. Res. Symp.
Proc. 313 (1993) 685
\end{thebibliography}
\end{document}